\documentclass[12pt]{article}
\usepackage[pdftex]{graphicx}
\usepackage{hyperref}
\usepackage{amsmath}
\usepackage{amssymb}
\renewcommand{\title}[1]{%
    \bigskip%
    \begin{center}%
    \Large\bf #1%
    \end{center}%
    \vskip .2in}

\renewcommand{\author}[1]{%
    {\begin{center}
    #1
    \end{center}}}

\begin{document}

\title{\bf{Covariant Formulation  of the Newton-Hooke Particle and its Canonical Analysis}}

\centerline{{Rabin Banerjee}\footnote {rabin@bose.res.in}}
\bigskip

\centerline{S. N. Bose National Centre 
for Basic Sciences
, JD Block, Sector III, Salt Lake City,} 
\centerline{Kolkata -700 106, India }

\vskip 1cm
\begin{abstract}
\noindent A covariant formulation for the Newton-Hooke particle is presented by following an algorithm developed by us \cite{BMM1,
BMM2, BMM3}. It naturally leads to a coupling with the Newton-Cartan geometry. From this result we provide an understanding of gravitation in a Newtonian geometric background. Using Dirac's constrained analysis a canonical formulation for the Newton-Hooke covariant action is done in both gauge independent and gauge fixed approaches. While the former helps in identifying the various symmetries of the model, the latter is able to define the physical variables. From this analysis a  path to canonical quantisation is traced and the Schroedinger equation is derived which is shown to satisfy various consistency checks. Some consequences of this equation are briefly mentioned.
\end{abstract}

\section{Introduction }

\smallskip
The possibility of non-relativistic  symmetries  in a gravitational background was perhaps first  noticed by Cartan \cite{Cartan-1923},\cite {Cartan-1924},
who developed a covariant geometrical theory of Newtonian gravitation some time after Einstein formulated his general theory of relativity. The corresponding Newton-Cartan (NC) manifold has a degenerate metric structure and  the elements of NC geometry are used to couple the matter sector, be it particles \cite{Kuch, PPP, BCG, BMN1, banmukh}, extended objects or fields \cite{BGMM, j1}, with non-relativistic (NR) gravity. This geometrical approach has brought a resurgence in this field of research leading to various applications in condensed matter systems, hydrodynamics, particle physics and cosmology.

Due to the presence of degenerate metric structures, there is no straightforward method, contrary to the relativistic case, of coupling NR matter with gravity. In this context we have developed a structured algorithm in a set of papers \cite{BMM1, BMM2, BMM3,RB}. It is based on localising the non-relativistic (galilean) symmetry and naturally leads to a covariant formulation with a  geometric interpretation involving the NC structures \cite{BMM2}. Named as galilean gauge theory \cite{BM4}, it has found applications in different contexts, reproducing familiar results and also yielding new findings and insights \cite{BM4, BM5, BM6, BM7, MS,  BMN2}.

In this paper we apply our methodology to the Newton-Hooke theory. It is different from the previous examples since it is a NR theory in non flat space time. As such, this theory is useful for studying non-relativistic cosmological models with a cosmological constant \cite{ABCP, GP, china, AM} and has interesting connections with other branches of physics \cite{duval, bzy}. It is physically equivalent to the harmonic oscillator (or its inverted version) depending on whether the cosmological constant is positive or negative. A covariant formulation of this theory is desirable. We find that localising the symmetry of the usual Newton Hooke theory, a covariant form for the action is obtained where the coupling with the Newton- Cartan geometry is naturally revealed. Several aspects of this theory are investigated, including an obtention of the Schroedinger equation.

In section 2, we discuss the basics of Newton-Hooke groups. The algebra of these groups is obtained as a contraction of the de Sitter groups. While this is known \cite{ABCP, china} , we have shown this in a simple way that includes central extension. Also, some of this material is used in later sections. The representations for the group generators are given in both kinematic (purely algebraic) and dynamical terms. Then, in section 3,  the action for the Newton-Hooke theory is derived as a non-relativistic limit of the relativistic free particle moving in the de Sitter background. The limiting prescription is identified with that of the contraction process carried out in section 2. A detailed study of the Newton-Hooke symmetries is done and a representation of the generators, with or without central extension, is derived. Here we also show that, by an appropriate interpretation, the Newton-Hooke theory  admits the galilean symmetry, besides the usual Newton-Hooke symmetry. Section 4 briefly introduces the elements of our methodology, called galilean gauge theory. Its application to the Newton-Hooke theory is done in section 5 where the model in curved background is seen to involve a coupling with  Newton-Cartan geometry. In section 6 a lagrangian analysis is presented which helps in establishing a connection of the geometric formulation of Newtonian gravity with the present work. A hamiltonian formulation of the covariant Newton-Hooke  theory is given in section 7 employing both gauge independent and gauge fixed (section 7.1) approaches. The Schroedinger equation is derived in section 7.2 which also includes certain consistency checks. Finally, our conclusions are presented in section 8.

\section{Basics of Newton-Hooke Groups}

 There are two Newton-Hooke groups which presumably appeared originally in the paper \cite{BLL}. These were  further explored in \cite{DD} and have an important role in formulating non-relativistic cosmological models \cite{ABCP, GP, china, AM}. They are connected to the two de Sitter groups $SO(3, 2)$ and $S(4, 1)$ in the same way as the galilean group is connected to the Poincare group. Likewise, the de Sitter and Poincare groups are related just as their nonrelativistic counterparts, Newton-Hooke and galilean groups, are. The various connections are usually studied by means of contraction or deformation processes. In the case of  contraction, the resulting algebra is the one where some nonvanishing structure constants of the original group algebra vanish. For deformations, it is the other way round. Thus, the passage from the Poincare to the galilean or from the  de Sitter to the Newton-Hooke is a contraction. On the other hand, the passage from the galilean to the Poincare or to the Newton-Hooke are deformations. Incidentally, these are the two deformations of the galilean group.
 
 Here we study the contraction of the de Sitter groups to the two Newton Hooke groups.  The explicit forms for the generators will be obtained. Representations in specific coordinates will be given. These will be subsequently used to define the Newton Hooke particle in a NR curved background that naturally yields the coupling with the Newton Cartan geometry.
 
 The de Sitter spacetime can be defined as a hyperboloid of radius $R$ embedded in five dimensional flat (pseudo-Euclidean) spacetime whose coordinates satisfy the condition,
 \begin{equation}
 x^2 = \eta_{AB}x^A x^B = \epsilon R^2; \,\,\, \eta_{AB}= diag.( -1, +1, +1, +1,\epsilon); \,\, A,B= 0,1,2,3,4= \mu, 4
 \label{desitter}
 \end{equation}
where $\epsilon = +1 (-1)$ correspond, respectively, to the de Sitter (anti-de Sitter) cases. The de Sitter group of transformations of the five dimensional spacetime, leaving the hyperboloid invariant, can be identified with the five dimensional Lorentz group whose ten generators $J_{AB}$ satisfy the algebra,
\begin{equation}
[J_{AB}, J_{CD}] = \eta_{AC} J_{BD} - \eta_{AD} J_{BC} + \eta_{BD} J_{AC} - \eta_{BC} J_{AD}
\label{lorentz}
\end{equation}
The interpretation of these generators as the generators of the de Sitter group is effected by splitting them into the six generators of the four dimensional Lorentz group $J_{\mu\nu}$, satisfying an algebra identical to (\ref{lorentz}), plus four translation generators defined as, $P_\mu= \frac{1}{cR}J_{4\mu}$, where $c$ is the velocity of light. The complete de Sitter algebra is then obtained from (\ref{lorentz}) as,
\begin{eqnarray}
[J_{\mu\nu}, J_{\lambda\rho}] &=& \eta_{\mu\lambda} J_{\nu\rho} - \eta_{\mu\rho} J_{\nu\lambda} + \eta_{\nu\rho} J_{\mu\lambda} - \eta_{\nu\lambda} J_{\mu\rho}\cr
[J_{\mu\nu}, P_\lambda]&=& \eta_{\mu\lambda} P_\nu - \eta_{\nu\lambda} P_\mu  \cr
[P_\mu, P_\mu] &=& \frac{1}{c^2 R^2}[J_{4\mu}, J_{4\nu}]=\Lambda J_{\mu\nu} 
\label{desitteralgebra}
\end{eqnarray}
where $\Lambda= \frac{\epsilon}{c^2 R^2}$ is the cosmological constant which is positive (negative) for de Sitter (anti-de Sitter) spcetimes depending on the signature of $\epsilon$. The passage to the flat limit is obtained by setting the cosmological radius to infinity when the de Sitter four momentum
no longer exists and we recover the usual Lorentz symmetry.

In order to get the Newton Hooke algebra it is useful to express the de Sitter algebra in terms of the translation generators $P_i$,  redefined angular momentum generators $J_i$, boosts $K_i$ and the hamiltonian $H$,
\begin{equation}
J_i = \frac{1}{2}\epsilon_{ijk}J_{jk} ;\,\,\,  K_i=J_{0i};\,\,\,  H=P_0
\label{newgenerators}
\end{equation}
where the Latin indices run from $1$ to $3$. Then the de Sitter algebra has the structure,
\begin{eqnarray}
[J_i, J_j] &=& \epsilon_{ijk}J_k \,\,\, \,\,\,\,\, \,\,[J_i, K_j]= \epsilon_{ijk}K_k \,\,\,\,\,\,  [J_i, P_j]= \epsilon_{ijk}P_k \cr
[P_i, P_j] &=&  \Lambda \epsilon_{ijk} J_k\,\,\, [K_i, K_j]= - \epsilon_{ijk} J_k \,\,\, [K_i, P_j]= -\eta_{ij}H \cr
[H, P_i] &=&  \Lambda K_i\,\,\,\,\,\,\,\,\,\,\, [H, J_i]=0\,\,\,\,\,\,\, [H, K_i]= P_i
\label{finaldesitteralgebra}
\end{eqnarray}

To contract the above algebra to the Newton Hooke, the following generators, which involve the $0$-components,  have to be scaled,
\begin{equation}
P_0 :\,\,\, H^{NH} = c H^{DS}\,\,;\,\,\, J_{0i}:\,\,\,  K_i^{NH} = \frac{K_i^{DS}}{c}
\label{scaling}
\end{equation}
The Newton Hooke algebra follows from (\ref{finaldesitteralgebra}) with the above substitution followed by the limit $c\rightarrow \infty, \Lambda\rightarrow 0$ keeping $c^2\Lambda$ fixed leading to the  brackets,
\begin{eqnarray}
[J_i, J_j] &=& \epsilon_{ijk}J_k \,\,\, \,\,\,\,\, \,\,[J_i, K_j]= \epsilon_{ijk}K_k \,\,\,\,\,\,  [J_i, P_j]= \epsilon_{ijk}P_k \cr
[P_i, P_j] &=&  0 \,\,\,\,\,\,\,\,\,\,\, [K_i, K_j]= 0 \,\,\, \,\,\,\,\,\,\,\,[K_i, P_j]= 0\cr
[H, P_i] &=&  \frac{\epsilon}{R^2} K_i\,\,\,\,\,\,\,\,\,\,\, [H, J_i]=0\,\,\,\,\,\,\, [H, K_i]= P_i
\label{newtonhookealgebra}
\end{eqnarray}

If the zero point term is included by redefining the hamiltonian as
 $H^{NH}= cH^{DS}- mc^2$ and then the contraction is performed, we obtain the centrally extended Newton Hooke algebra, where the $K-P$ bracket gets replaced by,
\begin{equation}
[K_i, P_j]= -m\eta_{ij}
\label{ctnh} 
\end{equation}
If the flat limit $R\rightarrow \infty$ is taken,  then the Newton Hooke algebra contracts to the galilean algebra, with or without central extension,
\begin{eqnarray}
[J_i, J_j] &=& \epsilon_{ijk}J_k \,\,\, \,\,\,\,\, \,\,[J_i, K_j]= \epsilon_{ijk}K_k \,\,\,\,\,\,  [J_i, P_j]= \epsilon_{ijk}P_k \cr
[P_i, P_j] &=&  0 \,\,\,\,\,\,\,\,\,\,\, [K_i, K_j]= 0 \,\,\, \,\,\,\,\,\,\,\,[K_i, P_j]= 0 \,\, or \,\,[K_i, P_j]= -m\eta_{ij} \cr
[H, P_i] &=&  0 \,\,\,\,\,\,\,\,\,\,\, [H, J_i]=0\,\,\,\,\,\,\, [H, K_i]= P_i
\label{galileanalgebra}
\end{eqnarray}
\subsection{Representations}

A suitable representation for the de Sitter groups may be obtained by noting that the translation generators do not commute due to the presence of a curvature, characterised by the cosmological constant $\Lambda$. Thus its canonical form has to be appropriately modified to include such a term. One such choice is given by,
\begin{equation}
P_\mu= -(\partial_\mu + \Lambda x_\mu x^\lambda\partial_\lambda)
\label{translation}
\end{equation}
which is inspired from the Beltrami coordinates used to define the de Sitter space \cite{Mignemi}, reproducing the expected non-canonical form for the algebra,
\begin{equation}
[x_\mu, P_\nu]= \eta_{\mu\nu} + \Lambda x_\mu x_\nu
\label{beltrami}
\end{equation}
Taking the usual form of the Lorentz generators,
\begin{equation}
J_{\mu\nu} = x_\mu P_\nu - x_\nu P_\mu = - x_\mu \partial_\nu + x_\nu \partial_\mu
\label{lorentz1}
\end{equation}
with $P_\mu$ defined as (\ref{translation}) it is easy to verify the complete de Sitter algebra (\ref{desitteralgebra}).

In this representation the boosts and angular momentum retain their usual (flat background) expressions,
\begin{equation}
J_i= \epsilon_{ijk} x_k \partial_j \,\,\, ; \,\,\,\, K_i= x_i\partial_0 - x_0 \partial _i
\label{flatgenerators}
\end{equation}
while the translation generators (\ref{translation}) are modified.

Besides the dynamic representation discussed above there is a simple kinematic representation which uses the elements of a Clifford algebra $\gamma_A$ defined as,
\begin{equation}
\{\gamma_A, \gamma_B \}= -2 \eta_{AB}
\label{clifford}
\end{equation}
where the metric was introduced  in (\ref{desitter}). Then the Lorentz and translation generators for the de Sitter groups are given by,
\begin{equation}
J_{\mu\nu} = \frac{1}{4} [\gamma_\mu, \gamma_\nu] \,\,\,\,;\,\,\,\,  P_\mu= \frac{1}{2cR}\gamma_\mu \gamma_4
\label{clifford1}
\end{equation}

For the two Newton Hooke groups following from the two de Sitter groups by contraction, a representation for the generators is given in terms of circular functions for the  case $(\epsilon= -1)$ and hyperbolic functions for the  case $(\epsilon=+1)$. Specialising to the former, one finds \cite{GP},
\begin{equation}
H=\partial_t \,\,\,,\,\,\, P_i=-\cos(\frac{t}{R})\partial_i\,\,\,, \,\, K_i=-R \sin(\frac{t}{R}) \partial_i\,\,\,, \,\,\, J_i=\epsilon_{ijk} x^k\partial_j
\label{dsgenerators}
\end{equation}
Subsequently, in section 3, we provide a systematic derivation of these expressions which also generalises to the case of central extension. By replacing $t\rightarrow it$ the results for the $\epsilon=+1$ case are obtained. Setting the cosmological radius $R\rightarrow\infty$ yields the usual galilean generators,
 \begin{equation}
H=\partial_t \,\,\,,\,\,\, P_i=-\partial_i\,\,\,, \,\, K_i=-t\partial_i\,\,\,, \,\,\, J_i=\epsilon_{ijk} x^k\partial_j
\label{galgenerators}
\end{equation}

The difference in the structures (\ref{dsgenerators}) and (\ref{galgenerators}) is intimately connected with translation and boost invariances. The fact that, in the galilean case, the boosts have an explicit time dependence while the translations do not is due to the different algebra for the $H-P$ and $H-K$ brackets (\ref{galileanalgebra})that enforce the conservation of the respective generators,
\begin{eqnarray}
\frac{dP_i}{dt}&=& \frac{\partial P_i}{\partial t} + [P_i, H] = 0\cr
\frac{dK_i}{dt}&=& \frac{\partial K_i}{\partial t} + [K_i, H] = P_i - P_i =0
\label{consistency}
\end{eqnarray}

 The fundamental difference of the Newton Hooke algebra (\ref{newtonhookealgebra}) from the galilean is the nontrivial $H-P$
  bracket.  This is necessary to ensure the conservation of the translations,
  \begin{equation}
  \frac{dP_i}{dt}= \frac{\partial P_i}{\partial t} + [P_i, H] = \frac{1}{R} \sin(\frac{t}{R})\partial_i + \frac{K_i}{R^2} =0
  \label{consistency1}
  \end{equation}
  Similarly the boosts $K_i$ are also conserved. 

As we shall show there are other representations. Specifically we elaborate on an example that is physically motivated. To do this we first derive an action for the Newton Hooke particle.

\section{Action for Newton Hooke Particle and its Symmetries}

Just as the de Sitter groups could be contracted to the Newton Hooke groups using a specific prescription it is possible to obtain the action for the Newton Hooke particle by taking an appropriate non-relativistic limit of the relativistic point particle in the de Sitter background. This action is given by,
\begin{equation}
S=-mc\int d\tau \sqrt{-g_{\mu\nu} \dot x^\mu \dot x^\nu}
\label{freeparticle}
\end{equation}
where $m$ is the mass of the particle and the overdot denotes a derivative with respect to the parameter $\tau$. Here $g_{\mu\nu}$ is the de Sitter metric in global coordinates, defined as,
\begin{equation}
g_{00}= - f(r)\,\,\, , \,\, g_{rr}= \frac{1}{f(r)}\,\,\, , \,\,\, f(r)=1-\Lambda r^2 \,\,\, (r^2=x^ix^i)
\label{global}
\end{equation}
Expanding the square root and retaining  the terms that survive the limit $c\rightarrow \infty, \Lambda \rightarrow 0$ keeping $c^2\Lambda$ constant (which is the limit in which the de Sitter groups contracted to the Newton Hooke groups), one obtains,
\begin{equation}
S=  \frac{m}{2} \int d\tau\,\,\big(\frac{\dot x^i\dot x^i }{\dot t} + \frac{\epsilon \dot t x^i x^i}{R^2}\big) -mc^2 \int d\tau\,\, \dot t
\label{nhaction}
\end{equation}
where the cosmological constant has been expressed in terms of the cosmological radius. Thus, apart from a zero point energy term which is dropped from now onwards, the Newton Hooke action is the action for a simple harmonic oscillator. The potential will be repulsive (de Sitter) or attractive (anti- de Sitter)  depending on the sign of $\epsilon$.

The fact that the Newton Hooke action is just the harmonic
oscillator 
action allows us to find the symmetries in a simple manner. It is useful to recall that the space of motions of the harmonic oscillator is identical to the free particle \cite{DH}. The symmetries of the free particle are well known with the relevant generators, in the coordinate representation, given by,
\begin{equation}
H=\frac{p^2}{2m}\,\,\,;\,\, P_i=p_i\,\,\,;\,\, J_i=-\epsilon_{ijk} x_k p_j\,\,\, ;\,\,\, K_i= t p_i- m x_i\,\,\,
\,\, (p_i=-\partial_i)
\label{free}
\end{equation}
which satisfy the galilean algebra (\ref{galileanalgebra}) with a central extension.
The usual algebra without central extension  is obtained by taking the boost generator as $K_i=t p_i$.

An identical result for the galilean algebra, with or without the central extension, is obtained by taking the hamiltonian as 
$H=\partial_t$, instead of that given in (\ref{free}). In the case of the harmonic oscillator, discussed below, this leads to important differences.

 The hamiltonian of the oscillator corresponding to the lagrangian defined by the action in (\ref{nhaction}) (where we set the coordinate time equal to the parameter $(t=\tau)$) is given by,
 \begin{equation}
 H= \frac{p_i^2}{2m} + \frac{m x_i^2}{2R^2}
 \label{ham}
 \end{equation}
where we have taken the anti de Sitter case $(\epsilon=-1)$ and $p_i$ is the momenta conjugate to $x_i$. The classical trajectories are,
\begin{equation}
x_i(t)= a_i \cos(\frac{t}{R}) + (\frac{R}{m}) b_i \sin(\frac{t}{R})
\label{trajectory}
\end{equation}
where $a_i$ and $b_i$ parameterize the space of motions in  $   (\bf{R}^{2n})$ if we consider an $n$-dimensional oscillator. Now these parameters are expressed in terms of the phase space variables,
\begin{equation}
a_i= x_i \cos(\frac{t}{R})- (\frac{R}{m})p_i\sin(\frac{t}{R})\,\,\,\, ;\,\,\,\,b_i= (\frac{m}{R})x_i \sin(\frac{t}{R})+ p_i\cos(\frac{t}{R})
\label{solution}
\end{equation}
It can be checked that the space of motions is identical to the symplectic vector space of a free particle,
\begin{equation}
\bf{a}\wedge \bf{b}= \bf{x}\wedge\bf{p}\,\,\,\, ;\,\,\,\, [\bf{a}, \bf{b}]= [\bf{x}, \bf{p}]
\label{symplectic}
\end{equation}
 and hence carries a galilean symmetry \footnote{In fact the harmonic oscillator has the complete symmetry of the free particle, namely, the Schroedinger symmetry, that includes the conformal sector \cite{DH}. Here, however, we are interested only in the galilean sector.}.
 
 With this exercise it is possible to read off the relevant generators for the Newton Hooke theory by comparison with the free particle expressions. The results are,
 \begin{eqnarray}
 J_i&=& \epsilon_{ijk}a_j b_k= \epsilon_{ijk}x_j p_k\cr
 P_i &=& b_i = (\frac{m}{R})x_i \sin(\frac{t}{R})+ p_i\cos(\frac{t}{R})\cr
 H &=& \frac{b_i^2}{2m}= \frac{1}{2m} \Big((\frac{m}{R})x_i \sin(\frac{t}{R})+ p_i\cos(\frac{t}{R})\Big)^2\cr
 K_i&=& - m a_i= - m\Big ( x_i \cos(\frac{t}{R})- (\frac{R}{m})p_i\sin(\frac{t}{R})\Big )
 \label{galalgebra}
 \end{eqnarray}
which yields the centrally extended galilean algebra. 

It is possible to give a representation where the role of $H$ as a generator of time translation becomes manifest. In that case it is represented by,
\begin{equation}
    H= \partial_t - \frac{1}{2m}\Big(\sin^2\frac{t}{R}\,\,\partial^2 +\frac{m^2}{R^2}x^2 \cos^2\frac{t}{R} + \frac{m}{R} \sin \frac{2t}{R}\, x_i\partial_i\Big)
    \label{canham}
\end{equation}
while the other generators are simply obtained by replacing $p_i$ by $(-\partial_i)$ in (\ref{galalgebra}).

On the other hand if, instead of (\ref{canham}),  we take the usual Schroedinger representation for the hamiltonian $H=\partial_t$, keeping the other generators unchanged, then we obtain the centrally extended Newton Hooke algebra (\ref{newtonhookealgebra}) along with (\ref{ctnh}). The generators of the centrally extended Newton Hooke symmetry are now given by,
\begin{eqnarray}
 J_i&=& -\epsilon_{ijk}x_j \partial_k\cr
  H &=& \partial_t \cr
 P_i &=&(\frac{m}{R})x_i \sin(\frac{t}{R})- \cos(\frac{t}{R})\partial_i\cr
 K_i&=&- m\Big( x_i \cos(\frac{t}{R})+ (\frac{R}{m})\sin(\frac{t}{R})\partial_i\Big )
 \label{nhalgebra1}
 \end{eqnarray}

 The above representation for the centrally extended Newton Hooke generators complements the usual (non centrally extended) representation given in (\ref{dsgenerators}). By retaining only the derivative terms we reproduce (\ref{dsgenerators}). Thus, depending on the representation for the hamiltonian, the Newton Hooke theory conforms to either the galilean symmetry or the Newton Hooke symmetry. This is an important distinction from the free particle theory. It is simple to check that, in the flat limit $R\rightarrow\infty$, the centrally extended galilean representation (\ref{free}) is reproduced.
 
 The fact that the Newton-Hooke particle carries both the galilean and Newton-Hooke symmetries is useful for the curved space generalisation of this model. The point is that we have a definite prescription for obtaining the curved space result for any non-relativistic theory on a flat background that is galilean invariant. This is called galilean gauge theory which is briefly introduced in the next section. Thus we can safely extend our formalism to the Newton-Hooke particle by appropriately modifying our earlier analysis for the non-relativistic free particle.
 
 We now construct the covariant action of the Newton Hooke particle in a curved background. This will be done by exploiting the method of galilean gauge theory, advocated by us in a set of papers \cite{BMM1, BMM2, BMM3, RB}. As we shall see, this covariantisation naturally leads to a coupling with a background Newton-Cartan geometry. The elements of galilean gauge theory are now introduced.
 
 \section{Basics of Galilean Gauge Theory}

 Galilean gauge theory (GGT) is a technique that, given a non-relativistic theory in a flat background, is able to construct the corresponding theory in a curved background, following a structured algorithm. 
 
 The idea is to localise (or gauge) the original global non-relativistic symmetry by making the parameters of the transformation local (i.e. space time dependent).  Obviously the invariance of the action  is lost on gauging due to the presence of derivatives in the action. To recover the invariance under the local symmetry, ordinary derivatives are replaced by covariant derivatives, introducing new fields. The transformations of the new fields are fixed by demanding that, under local transformations, the covariant derivatives change in the same way as the ordinary ones did under the global transformations. This will naturally ensure the invariance of the new action  now written by replacing, in the old action, the ordinary derivatives by their covariant expressions. 
 
 Once the modified theory invariant under the local transformations has been obtained, it is possible to provide a geometrical interpretation. We can show that the new fields act as vielbeins(and their inverses) that connect the global coordinate basis with the local basis. Composites constructed from these vielbeins are seen to define the elements of the Newton-Cartan geometry.
 
 In earlier examples global galilean symmetry was gauged. Now we consider the Newton Hooke symmetry, which is more involved than the galilean since it contains time dependent circular functions (see, for instance, (\ref{nhalgebra1})).  Neverthelees, the basic algorithm goes through and a covariant form for the action is derived having similar geometrical interpretations. This is the purpose of the next section.
 
 \section{Covariantised Action for Newton Hooke Particle }
 
 The action (\ref{nhaction}) for the Newton Hooke particle is quasi invariant under the transformations,
 \begin{eqnarray}
 t&\rightarrow& t+q\cr
 x^i&\rightarrow& x^i + s^i \cos(\frac{t}{R}) + b^i R\sin(\frac{t}{R})+ \omega^{i}\,_jx^j\,\,; \,\, \omega_{ij}=\epsilon_{ijk}r_k
 \label{nhtrans}
 \end{eqnarray}
 where $q$ is the time translation parameter while $s_i, b_i, r_i$ respectively, are the space translation, boost and rotation parameters. These were obtained from the representation (\ref{dsgenerators}) corresponding to the noncentrally extended  Newton Hooke algebra. We could have also worked with the centrally extended version (\ref{nhalgebra1}) but the results would remain unchanged.

The parameters of the transformation are global, i.e. space time independent. We now gauge the symmetry by making these parameters space time dependent. Keeping in view the universal role of time, the time translation parameter $q$ becomes a function of only $t$ while the others are are functions of both $x$ and $t$, so that,
\begin{equation}
q\rightarrow q(t)\,\,\,;\,\,\, s_i, b_i, r_i \rightarrow f(x, t)
\label{functions}
\end{equation} 

Due to the presence of derivatives the action (\ref{nhaction}) is no longer invariant under the localised transformations. Galilean gauge theory spells out the method of obtaining the invariance. Under the global transformation (\ref{nhtrans}), the  derivatives transforms as,
\begin{equation}
\delta(\frac{dt}{d\tau})=\frac{d}{d\tau}(\delta t)=0\,\,;\,\, \delta\frac{dx^i}{d\tau}= -\frac{s^i}{R}\sin(\frac{t}{R})\frac{dt}{d\tau}+ b^i \cos(\frac{t}{R})\frac{dt}{d\tau} +\omega{^i}\,_j \frac{dx^j}{d\tau}
\label{globaltr1}
\end{equation}
Here $\tau$ is the parameter that appears in the integrand (\ref{nhaction}). In order to achieve invariance under the local transformations, the ordinary derivatives have to be first replaced by covariant derivatives. These covariant derivatives, under local transformations, are required to transform in the same way as ordinary derivatives did under global transformations. Then the new theory obtained by replacing ordinary derivatives in (\ref{nhaction}) by their covariant generalisations will be locally invariant. This is the principle of gauging. From (\ref{globaltr1}) the requisite transformations of the covariant derivatives are spelled out as,
\begin{equation}
\delta(\frac{Dt}{d\tau})=0\,\,\,;\,\,\,  \delta\frac{Dx^i}{d\tau}= -\frac{s^i}{R}\sin(\frac{t}{R})\frac{Dt}{d\tau}+ b^i \cos(\frac{t}{R})\frac{Dt}{d\tau} +\omega{^i}\,_j \frac{Dx^j}{d\tau}
\label{globaltr2}
\end{equation}
where the parameters are space time dependent according to (\ref{functions}). The next point is to define the covariant derivatives. Galilean gauge theory tells us the method. 
Following its tenets, we first observe that once the Newton Hooke symmetry is localised, it is necessary to introduce local coordinates at every point of space time that are connected with the global coordinates by,{\footnote{Indices from the beginning of the alphabet $(\alpha, \beta,..a,b,..)$ denote local coordinates while global ones are denoted from the middle of the alphabet $(\mu,\nu...i,j..)$. Here Greek indices indicate space time while Latin ones denote only space.}}
\begin{equation}
e^\alpha= \delta_\mu^\alpha e^\mu
\label{basisglobal}
\end{equation}
which is trivial at this point. Subsequently this will become nontrivial. 
The simplest possibility of defining the covariant derivative is a straightforward generalisation of (\ref{basisglobal}),
\begin{equation}
\frac{Dx^\alpha}{d\tau}= \Lambda{^\alpha}\,_\mu \frac{dx^\mu}{d\tau}\,\,\,;\,\, (x^0=t)
\label{covder}
\end{equation}
 The meaning of the new variable  $\Lambda$ will soon be clear. It is pointed out that the covariant derivative reduces to the ordinary derivative for the choice $\Lambda{^\alpha}\,_\mu =  \delta{^\alpha}\,_\mu$, mimicking the relation (\ref{basisglobal}). Using (\ref{globaltr2}) the transformation property of the new field is spelled out completely,
 \begin{equation}
 \delta \Lambda^a_\nu= -\partial_\nu T^\beta \Lambda^a_\beta + \omega^a_b \Lambda^b_\nu - b^a\Lambda^0_\nu\,\,\,;\,\, \delta \Lambda^0_0=0\,\,\,;\,\, \Lambda^0_i =0\,\,;\,\,(T^0= q \,\,, \,\,T^a= s^a)
 \label{vielbein}
 \end{equation}
It is now possible to provide a geometrical interpretation for the $\Lambda$ variables. From the above relations we observe that while its  local indices 
$(a)$ are Lorentz rotated, the global indices $(\nu)$ are coordinate transformed. The local Newton Hooke  transformation is thus interpreted as a non-relativistic  general coordinate transformation with $\Lambda^a_\nu$ playing the role of the inverse vielbein connecting local and global basis in a curved background. It is also noted that the same transformations were earlier found when gauging the galilean symmetry. The veilbein $\Sigma$ is defined as the inverse of $\Lambda$. These findings are now algebraically  expressed  as,
\begin{equation}
e_\mu=\Lambda^\alpha_\mu e_\alpha\,\,;\,\,e^\mu=\Sigma^\mu_\alpha e^\alpha\,\,\,;\,\, \Lambda^\alpha_\mu\Sigma^\mu_\beta=\delta^\alpha_\beta\,\\,;\,\, \Lambda^\alpha_\mu\Sigma^\nu_\alpha=\delta^\nu_\mu\,\,;\,\, e^\mu e_\mu= e^\alpha e_\alpha
\label{vielbeins}
\end{equation}

The gauging of the Newton Hooke symmetry has led to a curved space generalisation of the flat background. The flat space results are trivially obtained from (\ref{vielbeins}) by replacing the vielbeins by Kronecker deltas as the global and local bases become identical.

The following composites are now constructed from the vielbeins,
\begin{equation}
h^{\mu\nu}=\Sigma^\mu_a\Sigma^\nu_a\,\,;\,\, \tau_\mu=\Lambda^0_\mu=\Theta\delta^0_\mu \,\,;\,\, h_{\nu\rho}= \Lambda^a_\nu\Lambda^a_\rho\,\,;\,\, \tau^\mu=\Sigma^\mu_0
\label{nccomposites}
\end{equation} 
which, as shown earlier \cite{BMM2}, are the basic elements of the Newton-Cartan geometry that satisfy the Newton-Cartan algebra,
\begin{equation}
h^{\mu\nu}\tau_\nu=h_{\mu\nu}\tau^\mu=0\,\,;\,\, \tau^\mu \tau_\mu=1\,\,;\,\, h^{\mu\nu}h_{\nu\rho}= \delta^\mu_\rho - \tau^\mu \tau_\rho
\label{ncalgebra}
\end{equation}

The obtention of the Newton Hooke theory in a curved background now follows the standard algorithm. We first  replace the ordinary derivatives in the action (\ref{nhaction}) by the covariant derivatives. Then the first term in (\ref{nhaction}) simplifies to,
\begin{eqnarray}
\frac{Dx^a}{d\tau}\frac{Dx^a}{d\tau}&=&\Lambda^a_\mu \frac{dx^\mu}{d\tau}\Lambda^a_\nu \frac{dx^\nu}{d\tau}=h_{\mu\nu}\frac{dx^\mu}{d\tau}\frac{dx^\nu}{d\tau}\cr
\frac{Dx^0}{d\tau}&=&\Lambda^0_\mu \frac{dx^\mu}{d\tau}= \tau_\mu\frac{dx^\mu}{d\tau}
\label{curvedaction}
\end{eqnarray}
where we have used the relations (\ref{covder}) and (\ref{nccomposites}).

Similarly the quadratic piece in the second term in the action will be replaced by,
\begin{equation}
x^a x^a = (\Lambda ^a_\mu x^\mu) ( \Lambda^a_\nu x^\nu) = h_{\mu\nu} x^\mu x^\nu
\label{quadratic} 
\end{equation}
Hence, after these modifications, the original Newton Hooke action (\ref{nhaction}) (without the zero point term) takes the form,
\begin{equation}
S=  \frac{m}{2} \int d\tau\,\,\Big[h_{\mu\nu}\frac{dx^\mu}{d\tau}\frac{dx^\nu}{d\tau} \Big(\tau_\mu\frac{dx^\mu}{d\tau}\Big)^{-1} -\tau_\mu\frac{dx^\mu}{d\tau}\Big(\frac{ h_{\mu\nu}x^\mu x^\nu}{R^2}\Big)\Big] 
\label{nhncaction}
\end{equation}
According to our formalism this action will be (quasi) invariant under the gauged (local) Newton Hooke transformations. As expected, this covariant action, defined in a NR curved background, yields a geometrical interpretation. We find that the Newton Hooke particle now moves in a background Newton-Cartan spacetime. The coupling with the Newton-Cartan elements has come about naturally. The original action (\ref{nhaction}) (without the zero point term)  is easily recovered by passing to the flat limit $h_{0\mu}=0, h_{ij}=\delta_{ij}, \tau_0=1, \tau_i=0$.

\section{Lagrangian Analysis and Connection with Newtonian Gravity}

An analysis of the action (\ref{nhncaction}) will be done from which an insight into the geometric formulation of Newtonian gravity can  be gleaned. First, a quick review of Cartan's geometric formulation of Newton's theory is given. The laws of Newton are contained in the trajectory of neutral test particles,
\begin{equation}
    \frac{d^2 x^i}{dt^2} + \frac{\partial\Phi}{\partial x^i} =0
    \label{acc}
\end{equation}
where $x^i (i=1, 2, 3)$ are the spatial coordinates and the source equation for the potential $\Phi$ is given by,
\begin{equation}
    \nabla^2\Phi= 4\pi \rho G
    \label{source}
\end{equation}
where the mass density is denoted by $\rho$.  

In the conventional interpretation of Newton, the above equations describe a curved trajectory in three dimensional flat space. Cartan's covariantisation consists in interpreting the trajectories as geodesics (``straight lines") in four dimensional curved spacetime,
\begin{equation}
    \frac{d^2 x^\mu}{dt^2} + \Gamma^\mu_{\nu\rho}\frac{dx^\nu}{dt}\frac{dx^\rho}{dt}=0
    \label{geo}
\end{equation}
This is valid if one takes $x^\mu=(x^0=t, x^i)$ and also imposes the ansatz,
\begin{equation}
\Gamma^i_{00} = \frac{\partial\Phi}{\partial x^i}\,\,;\,\,\, \rm all \, \, other\,\, \Gamma^\mu_{\nu\rho}=0
\label{ansatz}
\end{equation}
so that (\ref{acc}) is reproduced. Also, inserting the ansatz in the usual expression for the Riemann tensor,
\begin{equation}
   R^\alpha_{\beta\gamma\delta}= \partial_\gamma \Gamma^\alpha_{\beta\delta}
   -\partial_\delta \Gamma^\alpha_{\beta\gamma} +\Gamma^\alpha_{\mu\gamma}\Gamma^\mu_{\beta\delta} -\Gamma^\alpha_{\mu\delta}\Gamma^\mu_{\beta\gamma}
   \label{riemann}
\end{equation}
one finds the only non-vanishing component to be given by,
\begin{equation}
    R^i_{0j0} = \partial_i\partial_j\Phi
    \label{riemann1}
\end{equation}
so that Poisson's equation (\ref{source}) is expressed as,
\begin{equation}
R_{00}= 4\pi\rho G
\label{poisson2}
\end{equation}
The geometric formulation of Newton's gravity is summarised in the above set of equations (\ref{geo}) to (\ref{poisson2}).

In order to make a contact  with our work, we now derive the Lagrange's equations of motion to put it in the form of the geodesic equation (\ref{geo}) from which the non-relativistic connection can be identified. The Euler-Lagrange equations,
\begin{equation}
    \frac{d}{d\tau}\Big(\frac{\partial L}{\partial x'^\mu}\Big)-\frac{\partial L}{\partial x^\mu}=0
    \label{el}
\end{equation}
are computed from the lagrangian defined in (\ref{nhncaction}), where a prime indicates derivative with respect to $\tau$. The basic equation obtained from (\ref{el}) is contracted by $h^{\gamma\mu}$ and relations (\ref{ncalgebra}) are used which considerably simplify the algebra leading to the result,
\begin{eqnarray}
   x''^{\gamma}&+& (\tau'.x')\tau^\gamma - \frac{(\tau.x')'}{(\tau.x')} x'^\gamma -\frac{h^{\gamma\alpha}}{2\tau.x'}h_{\rho\beta}x'^\rho x'^\beta(\tau'_\alpha - (\partial_\alpha\tau_\sigma) x'^\sigma)
    + \frac{h^{\gamma\alpha}}{2}(\partial_\sigma h_{\alpha\beta} +\partial_\beta h_{\alpha\sigma} - \cr
    &-&\partial_\alpha h_{\sigma\beta})x'^\sigma x'^\beta
    +\frac{\tau.x'}{2R^2}\Big(h^{\gamma\mu} \partial_\mu h_{\alpha\beta} x^\alpha x^\beta + 2 x^\gamma - 2\tau^\gamma (\tau.x)\Big)=0
    \label{big}
\end{eqnarray}
where the notation,
\begin{equation}
    \tau.x= \tau_\mu x^\mu
    \label{notation}
\end{equation}
has been used. Moreover, using the identity,{\footnote{The second equality in (\ref{identity}) follows since we are dealing with torsionless Newton-Cartan theory and the factor appearing in the parentheses is just the temporal component of the torsion.}}
\begin{equation}
    \tau'_\alpha - (\partial_\alpha\tau_\sigma) x'^\sigma= (\partial_\sigma\tau_\alpha-\partial_\alpha\tau_\sigma)x'^\sigma=0
    \label{identity}
\end{equation}
the fourth  term in (\ref{big}) can be dropped. The Dautcourt connection is next introduced,
\begin{equation}
    \Gamma^\gamma_{\sigma\beta}= \frac{\tau^\gamma}{2} (\partial_\sigma \tau_\beta + \partial_\beta \tau_\sigma)
    + \frac{h^{\gamma\alpha}}{2}(\partial_\sigma h_{\alpha\beta} +\partial_\beta h_{\alpha\sigma} - \partial_\alpha h_{\sigma\beta})+\frac{1}{2} h^{\gamma\alpha}(K_{\alpha\sigma}\tau_\beta + K_{\alpha\beta} \tau_\sigma)
    \label{daut}
\end{equation}
where $K_{\alpha\beta}$ is an arbitrary two form. 
Then (\ref{big}) can be written in terms of the Dautcourt connection as,
\begin{equation}
    x''^\gamma +\Gamma^\gamma_{\sigma\beta} x'^\sigma x'^\beta = \frac{(\tau.x')'}{\tau.x'}x'^\gamma  +h^{\gamma\alpha}K_{\alpha\sigma}\tau_\beta x'^\sigma x'^\beta - \frac{\tau.x'}{2R^2}\Big(h^{\gamma\mu} \partial_\mu h_{\alpha\beta} x^\alpha x^\beta + 2 x^\gamma - 2\tau^\gamma (\tau.x)\Big)
    \label{biggeo}
\end{equation}
While the left side has the expected terms of the geodesic equation, the other one does not. This is to be expected due to the presence of the potential (cosmological)  term. For free motion this has to be set to zero which is achieved by taking the large cosmological radius limit $(R\rightarrow \infty)$. Further, we observe that there are no suitable terms that can be absorbed in the arbitrary two form. Thus we take $K_{\alpha\beta}=0$. This yields the geodesic equation for free motion,
\begin{equation}
    x''^\gamma +\Gamma^\gamma_{\sigma\beta} x'^\sigma x'^\beta = \frac{(\tau.x')'}{\tau.x'}x'^\gamma
    \label{geofree}
\end{equation}

To obtain the standard (affine) form of the geodesic equation (\ref{geo}), we have to identify the affine parameter. The affine properties of Newton-Cartan geometry are determined by the direction of flow of time. The Galilean frame assumed here orients the time axis along the direction of absolute time. Substituting $\tau=t$ in the above equation yields,
\begin{equation}
    \frac{d^2x^\gamma}{dt^2} + \Gamma^\gamma_{\sigma\beta}  \frac{dx^\sigma}{dt}\frac{dx^\beta}{dt} =\frac{\dot \Theta}{\Theta}\frac{dx^\gamma}{dt}
\end{equation}
where an overdot denotes time $(t)$ differentiation.  The affine parameter fixing the scale of time is now defined as,
\begin{equation}
    dT=\Theta dt
    \label{affine}
\end{equation}
which leads to the usual affine geodesic equation,
\begin{equation}
    \frac{d^2x^\gamma}{dT^2} + \Gamma^\gamma_{\sigma\beta}  \frac{dx^\sigma}{dT}\frac{dx^\beta}{dT} = 0
    \label{affinegeo}
\end{equation}
and reproduces (\ref{geo}).

It is now possible to exactly map the Newtonian potential with the elements of Newton-Cartan geometry defined in (\ref{ncalgebra}). The independent components of these elements are first computed. The two vectors $(\tau_\mu, \tau^\mu)$ together have 8 components. Similarly, the two symmetric tensors $(h_{\mu\nu}, h^{\mu\nu})$ have 20 components. This amounts to 28 components. Now the relations (\ref{ncalgebra}) imply 25 conditions so that there are 3 independent elements. In the adapted (Galilean) coordinates where a three dimensional slicing is done at each instant of time, the Newton-Cartan elements are given by,
\begin{equation}
 \tau_\mu=\partial_\mu t=\delta^0_\mu, \, \tau^0=1,\, \tau^i=\tau^i,\, h^{ij}=h_{ij}=\delta_{ij},\, h^{0i}=0,\, h_{00}=\tau^i\tau^i,\, h_{0i}= -\tau^i 
 \label{adapted}
\end{equation}

It is easy to see that the three independent components are $\tau^i$ and the parametrisation satisfies the algebra (\ref{ncalgebra}). The components of the connection, defined in (\ref{daut}) (with $K_{\mu\nu}=0$) are now evaluated. These are,
\begin{equation}
   \Gamma^0_{\mu\nu}=\Gamma^i_{kl}=0\,,\,
   \Gamma^i_{0k}=\frac{1}{2}
   (\partial_i\tau^k-\partial_k\tau^i)\,,\, \Gamma^i_{00}=-\dot\tau^i - \tau^j\partial_i\tau^j 
\end{equation}
Following the ansatz (\ref{ansatz}), $\Gamma^i_{0k}=0$ so that $\tau^i$ may be expressed as the gradient of a scalar,{\footnote{While the other connection components vanish on their own, $\Gamma^i_{0k}$ does not. This is related to the well known fact that the covariant formulation of Newton's gravity introduces additional structure, proportional to a coriolis force type term. Since such a term does not exist in Newton's gravity, this term must be dropped. In this sense Newton-Cartan gravity is more general than Newton's gravity.}}
\begin{equation}
    \tau^i=\partial^i\phi
    \label{scalar}
\end{equation}
With this choice,
\begin{equation}
    \Gamma^i_{00}=-\Big(\partial^i(\dot\phi +\frac{1}{2}(\partial^j\phi)^2)\Big)
    \label{last}
\end{equation}
From the ansatz (\ref{ansatz}) and (\ref{last}), the  relation between the Newtonian potential and the Newton-Cartan element follows immediately,
\begin{equation}
    \Phi= -\Big(\dot\phi +\frac{1}{2}(\partial^j\phi)^2\Big)\label{map}
\end{equation}
and Newton's law of gravity is expressed with the help of the above equations and (\ref{poisson2}) as,
\begin{equation}
R_{00}=-\nabla^2(\dot\phi +\frac{1}{2}(\partial^j\phi)^2) = 4\pi\rho G    
\end{equation}
This shows how Newton's formulation and the geometric formulation of Cartan are related, using our results. 

The hamiltonian formulation is next developed as a sequel to the lagrangian analysis.

\section{Hamiltonian Formulation}

A detailed canonical analysis of the action (\ref{nhncaction}) will be performed eventually leading to the form of the Schroedinger equation. The canonical momenta are defined by,
\begin{equation}
p_\mu = \frac{\partial L}{\partial x'^\mu}= m\frac{h_{\mu\nu}x'^\nu}{\tau.x'}- m\frac{h_{\rho\nu}x'^\rho x'^\nu}{2(\tau.x')^2}\tau_\mu- m\frac{h_{\rho\nu}x^\rho x^\nu}{2R^2}\tau_\mu
\label{momenta}
\end{equation}
where the prime denotes a differentiation with respect to the parameter $\tau$ and a shorthand notation is introduced,
\begin{equation}
\tau.x'=\tau_\alpha x'^\alpha
\label{shorthand}
\end{equation}
From (\ref{momenta}) and the Newton-Cartan algebra (\ref{ncalgebra}) it follows,
\begin{eqnarray}
\tau^\mu p_\mu &=&-\frac{m}{2}\Big(\frac{h_{\rho\nu}x'^\rho x'^\nu}{(\tau.x')^2}+ \frac{h_{\rho\nu}x^\rho x^\nu}{R^2}\Big)\cr
h^{\alpha\mu} p_\mu &=& m\frac{x'^\alpha- \tau^\alpha \tau.x'}{\tau.x'}\cr
h^{\alpha\mu} p_\mu p_\alpha &=& m^2\frac{h_{\alpha\beta}x'^\alpha x'^\beta}{(\tau.x')^2}\label{momentarelations}
\end{eqnarray}

Using these relations we find the existence of a constraint,
\begin{equation}
\Phi_1= \tau^\mu p_\mu + \frac{h^{\alpha\mu} p_\mu p_\alpha}{2m} + m\frac{ h_{\rho\nu} x^\rho x^\nu}{2R^2} \approx 0
\label{constraint}
\end{equation}
which is implemented weakly in the sense of Dirac's approach \cite{dirac}. To check whether there are more constraints we have to verify its time conservation. This is done by first finding the canonical hamiltonian,
\begin{equation}
H_c= p_\mu x'^\mu - L = 0
\label{hamiltonian}
\end{equation}
which, expectedly, vanishes on account of the reparametrisation symmetry of the model. The entire dynamics is therefore governed by the constraint which enters into the picture through the total hamiltonian,
\begin{equation}
H_t = H_c + \chi \Phi_1
\label{total}
\end{equation}
where $\chi$ is a lagrange multiplier enforcing the constraint. Since $\Phi_1$ is strongly involutive,
\begin{equation}
\{\Phi_1, \Phi_1\} =0
\label{involutive}
\end{equation}
it is the only constraint. Also, it is first class and acts as the generator of the reparametrisation symmetry, as will soon be shown.

The lagrange multiplier $\chi$ may be determined by demanding consistency with the equation of motion,
\begin{equation}
    x'^\mu = \{x^\mu, H_t\}_{PB}= \chi\Big(\frac{x'^\mu}{\tau.x'}\Big)
    \label{multipliereqn}
\end{equation}
so that,
\begin{equation}
    \chi= \tau.x'
    \label{multiplier}
\end{equation}
and the total hamiltonian \cite{dirac} takes the form,
\begin{equation}
    H_t= (\tau.x')\Big(\tau^\mu p_\mu + \frac{h^{\alpha\mu} p_\mu p_\alpha}{2m} + m\frac{ h_{\rho\nu} x^\rho x^\nu}{2R^2}\Big)
    \label{totalham}
\end{equation}

If we take the $R\rightarrow \infty$ limit so that the original model is just the non-relativistic free particle in a curved background, then the above hamiltonian reduces to the super-hamiltonian given in \cite{Kuch}. We may thus interpret (\ref{totalham}) as the super-hamiltonian for the Newton Hooke particle in a curved background.

It is now possible to relate the reparameterisation symmetry with the gauge symmetry. Since there is only one first class constraint $\Phi_1$ the gauge generator is 
written as,
\begin{equation}
    G=\epsilon_1 \Phi_1= \epsilon_1 \Big(\tau^\mu p_\mu + \frac{h^{\alpha\mu} p_\mu p_\alpha}{2m} + m\frac{ h_{\rho\nu} x^\rho x^\nu}{2R^2}\Big)
    \label{gaugegenerator}
\end{equation}
where $\epsilon_1$ denotes the gauge parameter.

Then the gauge variation of the coordinates is given by,
\begin{equation}
    \delta x^\mu= \epsilon_1\{x^\mu, \Phi_1\}= \frac{\epsilon_1}{\tau.x'}x'^\mu
    \label{gaugevariation}
\end{equation}

Now the action (\ref{nhncaction}) has a manifest invariance under the finite  reparametrisation,
\begin{equation}
    \tau\rightarrow \tau'\,\,;\,\, x^\mu(\tau) \rightarrow x'^\mu(\tau')
    \label{reparametrisation}
\end{equation}
whose infinitesimal version is given by,
\begin{equation}
    \tau'=\tau + \delta \tau\,\,;\,\, \delta x^\mu(\tau) = \delta\tau \frac{dx^\mu}{d\tau}
    \label{infrep}
\end{equation}
Comparing with (\ref{gaugevariation}) we find the relation connecting the reparametrisation symmetry parameter with the gauge parameter,
\begin{equation}
    \delta \tau = \frac{\epsilon_1}{\tau.x'}
    \label{connection}
\end{equation}

One may obtain the hamiltonian equations of motion by Poisson bracketting the variables with the total hamiltonian. These are given as,
\begin{equation}
x'^\mu = \{x^\mu, H_t\}= (\tau.x')\Big(\tau^\mu + \frac{1}{m} h^{\mu\alpha}p_\alpha\Big) 
\label{e1}
\end{equation}
\begin{equation}
   p'_\mu = \{p_\mu, H_t\}= -(\tau.x')\Big(\partial_\mu\tau^\alpha p_\alpha + \frac{1}{2m} \partial_\mu h^{\rho\sigma} p_\rho p_\sigma + \frac{m}{R^2}h_{\mu\nu}x^\nu\Big) 
\label{e2} 
\end{equation}

and agree with the lagrangian equations of motion.

\subsection{Gauge fixing}

The above gauge independent analysis was useful for a study of the various symmetries and their connections. To identify the physical variables and develop the canonical quantisation, it is necessary to choose an appropriate gauge. A useful and, as seen later, physically viable gauge choice is to take the parameter $\tau$ identical to the universal time $x^0$ (or $t$),
\begin{equation}
    \Phi_2 = x^0 - \tau \approx 0
    \label{gauge}
\end{equation}

The constraints $\Phi_1$ and $\Phi_2$ are now second class and the matrix formed by the Poisson brackets of these constraints is given by,
\begin{equation}
    C_{ij}=\{\Phi_i, \Phi_j\}= -\Theta^{-1}\,\epsilon_{ij}\,\,;\,\, \epsilon_{12}=1\,\, (i, j=1,2)
\end{equation}
while its inverse is,
\begin{equation}
    C_{ij}^{-1}= \Theta \epsilon_{ij}
\end{equation}

The second class pair of constraints is now strongly implemented by working with the Dirac brackets (DB) instead of the Poisson. These brackets (denoted by a star) between any two variables are  given in terms of the Poisson brackets by,
\begin{equation}
    \{f, g\}^* = \{f, g\} - \{f, \Phi_i\} C_{ij}^{-1}\{\Phi_j, g\} 
\end{equation}
Then it is seen that the only relevant brackets, used to identify the canonical pair,  are given by,
\begin{equation}
    \{x^\mu, p_\nu\}^* = \delta^\mu_\nu - \Theta \delta^0_\nu \Big( \tau^\mu + \frac{1}{m} h^{\mu\sigma}p_\sigma\Big)
    \label{diracalgebra}
\end{equation}
The physical variables are now abstracted. Of the eight phase space variables, two are eliminated by the pair of constraints which are now strongly imposed. The remaining six physical degrees of freedom are given by the pair $(x^i, p_j)$. Indeed their DB are identical to their Poisson brackets, as seen from (\ref{diracalgebra}),
\begin{equation}
    \{x^i, p_j\} = \delta^i_j
    \label{canonicalpair}
\end{equation}
and hence these are the requisite canonical set that will be subsequently used for quantisation.

Once the physical set has been identified as $(x^i, p_j)$, the interpretation of the remaining pair $(x^0, p_0)$ becomes clear. The variable $x^0$ is just the time parameter and has vanishing DB with all variables since the constraint (\ref{gauge}) is now strongly imposed. The remaining variable $p_0$ is expressed in favour of the physical set by solving the constraint (\ref{constraint}),
\begin{equation}
    p_0 = -\Theta \Big( \tau^i p_i + \frac{1}{2m} h^{ij} p_i p_j + \frac{m}{2R^2} h_{\rho\nu}x^\rho x^\nu\Big)
    \label{newham}
\end{equation}

The last point is to define an appropriate hamiltonian since, after gauge fixing, the erstwhile total hamiltonian (\ref{totalham}), proportional to the constraint, vanishes. The new (gauge fixed) hamiltonian is given by,
\begin{equation}
    H= - p_0
    \label{gaugefixedham}
\end{equation}

This hamiltonian reproduces the equations of motion of the physical variables by appropriate Dirac bracketing,
\begin{equation}
  \dot x^i = \{x^i, H\}^*  \,\,;\,\, \dot p_i = \{p_i, H\}^*
  \label{finalequations}
\end{equation}
agreeing with the $i-th$ components of the respective equations (\ref{e1}) and (\ref{e2}), once the gauge condition (\ref{gauge}) is explicitly imposed.

\subsection{ Schroedinger equation from canonical quantisation}

Since the canonical variables have been found and, moreover, their Dirac brackets are identical to the Poisson brackets, it is straightforward to lift the Dirac algebra to commutators, replacing $x^i, p_i$ by operators, so that,
\begin{equation}
    [\hat x^i, \hat p_j] = i \delta^i_j
    \label{cmmutator}
\end{equation}
These operators have the standard forms in the coordinate representation,
\begin{equation}
    \hat x^i = x^i \,\,;\,\, p_i = -i\frac{\partial}{\partial x^i}
    \label{oprep}
\end{equation}

Using  (\ref{newham}), (\ref{gaugefixedham}) and (\ref{oprep}) we obtain,
\begin{equation}
    i\frac{\partial\psi}{\partial t} = \Theta\Big( \tau^i(-i \frac{\partial}{\partial x^i}) + \frac{1}{2m} h^{ij}(-\partial_i\partial_j) +\frac{m}{2R^2} h_{\rho\nu}x^\rho x^\nu\Big)\psi
    \label{sch}
\end{equation}

This is the Schroedinger equation for the Newton Hooke particle in a curved background where the  coupling with the Newton-Cartan geometry has appeared naturally. 

It is possible to make some consistency checks on the above result. If we take the flat limit in which case, $\Theta=1, \tau^i=0, h^{ij} = \delta^{ij} , h_{0\mu}=0, h_{ij}=\delta_{ij}$, the above equation simplifies to,
\begin{equation}
    i\frac{\partial\psi}{\partial t}= -\frac{1}{2m} \nabla^2\psi + \frac{m}{2R^2}x^2 \psi
    \label{flatsch}
\end{equation}
which reproduces the Newton Hooke Schroedinger equation in flat background or, alternatively, the equation for the harmonic oscillator (with $\omega^2= \frac{1}{R^2}$). If, further, the cosmological radius $R$ is set to infinity, we obtain the free particle Schroedinger equation. Likewise, the hamiltonian (\ref{newham}), in the same limits, first reduces to the result for the oscillator,
\begin{equation}
    H= \frac{p^2}{2m} + \frac{m}{2R^2} x^2
\end{equation}
and then to the expression for the free particle.

We now make some comments and observations related to the Schroedinger equation (\ref{sch}). This equation is written in the Newton-Cartan background. But, as discussed earlier, this has extra structure than is necessary for a covariant formulation of Newton's gravity. To write the Schroedinger equation in the background of Newtonian gravity, we make the identifications (\ref{adapted}) leading to the result,
\begin{equation}
    i\frac{\partial \psi}{\partial t} = \Big(\tau^i(-i\frac{\partial}{\partial x^i}) -\frac{1}{2m} \nabla^2 +\frac{m}{2R^2}(-2t\tau^i  x^i + \tau^i\tau^i t^2+ x^i x^i)\Big)\psi
    \label{nhsch}
\end{equation}
which naturally yields the same flat space limit (\ref{flatsch}).

The coupling of the Newton Hooke particle to other interactions, besides gravitation, is possible. To see this note that the Schroedinger equation (\ref{nhsch}) is invariant under the global $U(1)$ gauge transformations $\psi\rightarrow e^{i\theta}\psi$. To preserve this invariance under local transformations $\psi\rightarrow e^{i\theta (x, t)}\psi$ a gauge field is introduced transforming as $A_\mu\rightarrow A_\mu + i\partial_\mu\theta \,\, (\mu=t, i)$ and, as usual, ordinary derivatives in (\ref{nhsch}) are replaced by covariant derivatives, $\partial_\mu\rightarrow \partial_\mu - A_\mu$. The resulting Schroedinger equation describes the Newton Hooke particle minimally coupled to a background electromagnetic field.

It is also possible to check the self-adjointness of the operators. Since the above equation represents the hamiltonian it is self-adjoint. Self consistency is attained by interpreting $\tau^i$ as self-adjoint. From (\ref{scalar}) this implies that the scalar $\phi$ is real.

An interesting issue in classical dynamics is to include the effects of a variable mass. This is a practical problem and occurs in as simple and delightful a situation as the falling of raindrops or the much undesirable example of ballistic missiles and rockets. Traditionally, the effects of variable mass are included by suitably modifying Newton's second (force) law. In the present exercise, unfortunately, inclusion of variable mass poses severe problems. The point is that the constraint (\ref{constraint}) now has an explicit time dependence that enters through the mass term which is  written as $m(t)$ to show that it is no longer a constant. Its Poisson bracket with some other quantity is not known since we are working in the usual hamiltonian framework where Poisson brackets are defined at equal time and their quantisation leads to corresponding equal time commutators. Thus Poisson brackets of such explicitly time dependent objects are meaningless and we cannot proceed. Neither can we find the generator of gauge transformations (Gauss law) and hence gauge fixing cannot even be discussed. 
This is not to imply that the problem is undoable, rather it requires mathematical arsenal well beyond the scope of the present paper.{\footnote{For a hamiltonian treatment of time varying constraints, see \cite{time}. A lagrangian version is discussed in \cite{time1}.}}

\section{Conclusions}
We have derived an action for the Newton-Hooke particle in a curved background using our method of localising symmetries \cite{BMM1, BMM2, BMM3, RB}. This method, briefly reviewed here in section 4, was fruitfully applied earlier in different contexts. The final outcome was the coupling of the usual Newton-Hooke particle to a Newton-Cartan background, leading to a covariant form of the action.

The successful application of our method in this paper shows its generality and robustness since, in all previous examples, the  relevant global symmetry was galilean. In this case we localised the Newton-Hooke symmetries which involve either circular or hyperbolic functions. Nevertheless the basic  elements of Newton Cartan geometry were given by  exactly the same composites of the veilbeins (and their inverses) as was found earlier \cite{BMM2} for the galilean invariant models. Thus our method is general enough to discuss the coupling of gravity to non-relativistic symmetries that go beyond the usual galilean case.

An ab initio derivation of the representations for the generators of the Newton Hooke group, with or without central extension,  was given. Since the Newton Hooke action was physically equivalent to that of the harmonic oscillator (or the inverse oscillator), it was possible to use the mapping of the symmetries of the oscillator with those of the free particle, to get these representations. The Newton Hooke symmetries have a pivotal role in the subsequent analysis, which is based on localising them.

A lagrangian analysis was carried out. Specifically, the equations of motion in the zero torsion case were given and the Dautcourt connection was identified. The three independent elements of the Newton Cartan geometry were found that could be expressed as the gradient of a scalar. A direct connection of the equations of motion was established with Newton's equations, recast in a geometric form, by showing a map between this scalar and the gravitational potential. 

A detailed canonical analysis of the curved space covariant Newton Hooke action was performed. The first class constraint was identified which was shown to generate the reparameterisation (or the diffeomorphism) symmetry. It was possible to isolate the canonical (physical) variables of the theory by eliminating this freedom with an appropriate choice of gauge. This result was exploited to outline a quantisation program leading to the obtention of the Schroedinger equation. Some implications of this equation were briefly discussed.

As a final remark we mention that the covariant action found here involves only the elements of Newton Cartan geometry and there are no extraneous fields or variables. It could find applications in discussing  nonrelativistic cosmology, including possible effects of quantisation, in a covariant form.

\end{document}